\documentclass[12pt]{iopart}
\usepackage{iopams}
\usepackage{graphicx}

\begin{document}

\title[ ]{Neutron diffraction evidence for kinetic arrest of first-order
magneto-structural phase transitions in some functional magnetic materials}

\author{V Siruguri$^1$, P D Babu$^1$, S D Kaushik$^1$, Aniruddha Biswas$^2$, 
S K Sarkar$^2$, Madangopal Krishnan$^2$ and P Chaddah$^3$}

\address{$^1$UGC-DAE Consortium for Scientific Research Mumbai Centre, 
Bhabha Atomic Research Centre, Mumbai-400085, India}
\address{$^2$Glass and Advanced Materials Division, Bhabha Atomic Research 
Centre, Mumbai-400085, India}
\address{$^3$UGC-DAE Consortium for Scientific Research, University Campus, 
Khandwa Road, Indore-452001, India}

\ead{siruguri@csr.res.in}

\begin{abstract}

Neutron diffraction measurements, performed in presence of an external 
magnetic field, have been used to show structural evidence for the kinetic 
arrest of the first-order phase transition from (i) the high temperature 
austenite phase to the low temperature martensite phase in the magnetic
shape memory alloy Ni$_{37}$Co$_{11}$Mn$_{42.5}$ Sn$_{9.5}$, (ii) the higher 
temperature ferromagnetic phase to the lower temperature antiferromagnetic 
phase in the half-doped charge ordered compound La$_{0.5}$Ca$_{0.5}$MnO$_{3}$ 
and (iii) the formation of glass-like arrested states in both compounds. The 
cooling and heating under unequal fields protocol has been used to establish 
phase coexistence of metastable and equilibrium states, and also to demonstrate 
the devitrification of the arrested metastable states in the neutron diffraction 
patterns. We also explore the field-temperature dependent kinetic arrest line 
T$_{K}$(H), through the transformation of the arrested phase to the equilibrium 
phase. This transformation has been observed isothermally in reducing H, as also 
on warming in constant H. T$_{K}$ is seen to increase as H increases in both 
cases, consistent with the low-T equilibrium phase having lower magnetization.

\end{abstract}

\submitto{\JPCM}
\maketitle

\section{Introduction}

A number of functional magnetic materials like CMR manganites, magnetic shape 
memory alloys, intermetallics, multiferroics, etc., exhibit first order phase 
transitions (FOPT) as a function of temperature, pressure or magnetic field. 
While the microscopic nature of these FOPT has been extensively studied using 
temperature and pressure, the use of magnetic field as a useful thermodynamic 
variable in exploring these FOPT at an atomic scale is not very common. However,
in recent times, the study of first order transformation kinetics of these 
functional magnetic materials under the influence of a magnetic field has 
been subject of intense experimental research \cite{r1,r2,r3,r4,r5}. In 
these materials, typically, the high temperature magnetic phase can be
kinetically arrested when cooled under an appropriate external magnetic 
field, thereby inhibiting the transformation to the low temperature 
equilibrium phase. This kinetically arrested state has been shown to have 
coexisting phases of magnetically ordered metastable and equilibrium states
and such a material has been termed \cite{r6,r7} as a 'magnetic glass'. 
These studies attempt to investigate those regimes in knowledge space which 
have been hitherto unexplored, by using magnetic field as a useful 
thermodynamic variable. While the experimental techniques used in these 
studies are mostly based on transport and magnetization measurements, 
atomic scale measurements delineating the crystalline and magnetic 
structures as a function of magnetic field, for which neutron diffraction 
is a unique probe, have not been carried out so far. Certain classes of 
ferromagnetic shape memory alloys (FSMA) like Co-doped NiMnSn, NiMnIn and 
NiMnAl undergo a kinetic arrest of the first-order austenite to martensite
transition \cite{r6,r8,r9,r10}. In these materials, the higher temperature 
austenite phase has a higher ferromagnetic moment as compared to the lower 
temperature martensite phase and this manifests itself as a sharp drop in 
magnetization in a M versus T measurement. By increasing the magnetic field
of measurement, it has been observed that the kinetics associated with this 
first order transformation get hindered and beyond a critical field, they 
get completely arrested. Hence, there will be regions of field-temperature space 
where there will be a phase coexistence of metastable (arrested) or glass-like
arrested states (GLAS) and equilibrium (transformed) states. If the lower 
temperature equilibrium (transformed) state has a lower magnetization value 
than the high temperature austenite phase, it has been noted earlier that 
by cooling in a certain higher field (H$_{c}$) and warming in a lower
field (H$_{w}$) would lead to a de-arrest or devitrification of the 
GLAS \cite{r11}. On further warming, the devitrified state would undergo 
a reverse magnetic and structural transition to the high temperature,
high moment austenite phase. This novel protocol of cooling and heating 
in unequal fields (CHUF) \cite{r11} offers an unambiguous method to observe 
if the glass-like arrested state devitrifies, thereby qualifying it to be 
called a magnetic glass. While the transitions in this material are 
predominantly structural, kinetic arrest has also been observed in 
materials which undergo a ferromagnetic to antiferromagnetic transition, 
like, for example, the archetypal half-doped, charge-ordered manganite 
La$_{0.5}$Ca$_{0.5}$MnO$_{3}$ (LCMO) \cite{r12}. This compound is paramagnetic 
at room temperature, and upon cooling, it first enters a ferromagnetic (FM) 
state at $\sim  225$ K and then undergoes a FOPT into an antiferromagnetic 
(AFM) state at $\sim 150$ K \cite{r13}. Transport and magnetization 
measurements employing similar measurement protocols show that the 
ferromagnetic phase can be kinetically arrested when cooled under an 
appropriate magnetic  field through the FM-AFM transition, resulting in 
coexisting phases of metastable (arrested) FM states and equilibrium 
(transformed) AFM states \cite{r14}. The percentages of these phases 
would depend on the magnitude of the cooling field. CHUF protocol has 
been employed in the case of this compound also to demonstrate, through
magnetization measurements, the devitrification of these glass-like 
arrested metastable states \cite{r12}. While transport and magnetization 
measurements do not provide information about the structures at the 
microscopic scale, neutron diffraction is a powerful tool which can 
not only give information about the crystalline structure but also the 
magnetic structure. In particular, antiferromagnetic ordering of spins 
gives rise to unique reflections in neutron diffraction whose origin is 
purely magnetic in nature. Therefore, it is expected that neutron 
diffraction in presence of an external magnetic field, when carried out 
using the above measurement protocol, would give unambiguous information 
about coexisting phases, crystalline and magnetic, in a material. 

We present here neutron diffraction (ND) studies, in presence of an external 
magnetic field over a wide region of the field-temperature (H,T) space, on 
two widely different materials (i) the ferromagnetic shape memory (FSMA) 
alloy Ni$_{37}$Co$_{11}$Mn$_{42.5}$Sn$_{9.5}$ and (ii) half-doped manganite 
La$_{0.5}$Ca$_{0.5}$MnO$_{3}$. The former, on cooling, undergoes a FOPT from 
a high moment, austenite phase to a low moment, martensite phase while the 
latter compound undergoes a predominantly first order magnetic transition 
from a ferromagnetic state to an antiferromagnetic state on cooling. We 
show evidence, for the first time, for the kinetic arrest of the first 
order structural and magnetic phase transitions in these compounds. In 
Ni$_{37}$Co$_{11}$Mn$_{42.5}$Sn$_{9.5}$, the first order austenite to 
martensite transition gets kinetically arrested when cooled under an 
appropriate magnetic field, and on heating from 2 K to room temperature in 
a different field, a subsequent devitrification of the arrested metastable 
austenite state to the martensite state takes place, followed by the reentrant 
martensite-to-austenite transition. Similarly, in the half doped 
La$_{0.5}$Ca$_{0.5}$MnO$_{3}$, we show that the higher temperature 
ferromagnetic phase gets kinetically arrested on cooling under a magnetic 
field and when subsequently heated under a different field, the metastable 
arrested ferromagnetic state devitrifies into the stable antiferromagnetic 
state before reentering the higher temperature ferromagnetic state. The 
phase coexistence of the arrested metastable and equilibrium states by 
employing the CHUF protocol is demonstrated. Finally, we explain the (H,T) 
phase diagram for the two compounds using the T$_{K}$(H) line and attempt 
to draw analogies with the pressure-temperature (P,T) phase diagrams for 
metallic liquid germanium \cite{r15} and glassy water \cite{r16}. This study 
also brings out the versatility of magnetic field over hydrostatic pressure 
in that it is much easier to vary magnetic field as compared to pressure.

\section{Experimental details}

The Ni$_{37}$Co$_{11}$Mn$_{42.5}$Sn$_{9.5}$ buttons were
made by vacuum arc melting high purity (99.99\%) elements in appropriate
proportion. The buttons were solutionized at 1273 K for 24 h in sealed
quartz ampoule. High temperature Differential Scanning Calorimetry (DSC)
was used to confirm the two-phase melting behavior of this alloy and an
extended homogenization treatment was given to ensure structural and
chemical homogeneity. Characterization of this alloy has been carried
out using Optical and Scanning Electron Microscopy, Electron Probe
Microanalysis, XRD and neutron diffraction (ND), Transmission Electron
Microscopy (TEM), DSC and DC magnetization. Bulk cylindrical specimens
of 5 mm diameter were prepared for ND using electro-discharge machining,
as powder samples do not undergo martensitic transformation in this
alloy \cite{r5}. Nevertheless, one powder specimen was also prepared by
crushing a part of the treated button and checked for room temperature
structure. Specimens for TEM were prepared by slicing discs from an
electro-discharge machined cylindrical rod of 3 mm diameter, followed by
grinding and jet polishing with a Struers Tenupol-5 at 233 K, using a 10
vol \% perchloric acid in methanol electrolyte. For ND measurements, the
bulk cylindrical specimen was directly attached to the sample stick and
inserted into the cryomagnet. The La$_{0.5}$Ca$_{0.5}$MnO$_{3}$ sample was 
prepared by the conventional solid-state route and the single phase nature 
was confirmed using XRD whose pattern matched well with literature reports. 
Polycrystalline powder sample was compacted into pellets and inserted into 
the vanadium sample holders. This avoided any preferential orientations of 
the grains under a magnetic field. The ND patterns were collected using the 
position-sensitive detector based focusing crystal diffractometer installed 
by the UGC-DAE CSR Mumbai Centre \cite{r17} at the Dhruva reactor, Trombay, 
at a wavelength of 1.48 {\AA} in the temperature range of 2 K to 300 K under 
different conditions of applied magnetic field up to 7 Tesla. It was observed 
that the cylindrical FSMA sample consisted of extremely large grains which 
manifest themselves in almost single crystal-like behavior in the diffraction 
patterns. Hence, the sample was oriented such that two major reflections of 
the austenite phase at room temperature, namely, (111) and (200) were clearly 
visible in the diffraction patterns. The sample was locked in this orientation
for all subsequent measurements. DC magnetization was measured using a
commercial 9 T PPMS-VSM (make Quantum Design).

\section{Results and discussion}

\subsection{Ferromagnetic shape memory alloy Ni$_{37}$Co$_{11}$Mn$_{42.5}$Sn$_{9.5}$}
\subsubsection{Alloy characterization}

\begin{figure}
  \centering
   \includegraphics[width=8.00cm,height=6.5cm]{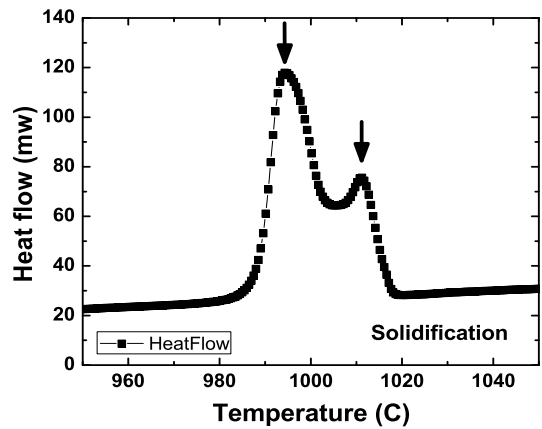}\\
  \caption{High temperature DSC plots of Ni$_{37}$Co$_{11}$Mn$_{42.5}$Sn$_{9.5}$ 
showing the two-phase incongruent melting behavior.}\label{fig1}
\end{figure}

\begin{figure}
  \centering
   \includegraphics[width=10.70cm,height=8.02cm]{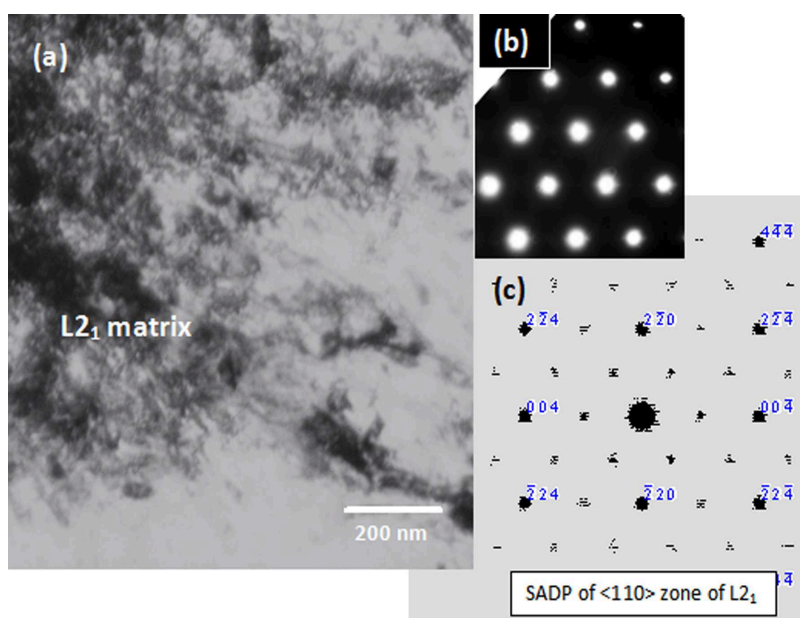}\\
  \caption{(a) Microstructure of the parent L2$_{1}$ phase in as-solutionized 
condition, (b) SAD pattern of the $<$110$>$ zone axis, and (c) its simulated key.}\label{fig2}
\end{figure}

\Fref{fig1} displays the high temperature DSC plots of this alloy. It 
clearly shows the two-phase incongruent melting behavior. Earlier, Watchel 
\etal \cite{r18} noted the presence of a peritectic reaction in the ternary 
Ni$_{0.5}$Mn$_{0.5-x}$Sn$_{x}$ ($0 \le x \le 0.5$) pseudo-binary phase 
diagram. More recently, Yuhasz \etal \cite{r19} also reported the problem 
of chemical and structural inhomogeneity that was prevalent in the ternary 
Ni-Mn-Sn alloys and was caused by this incongruent melting. It was thus, 
very important to solutionize these alloys for an extended period of time 
to ensure single-phase and uniform microstructure. Microstructure in 
as-solutionized condition shows the parent L2$_{1}$ phase (\fref{fig2}(a)), 
as confirmed by the $<$110$>$ SAD pattern and its simulated key (\fref{fig2}(b-c)).

\subsubsection{DC Magnetization}

\begin{figure}
  \centering
   \includegraphics[width=10.70cm,height=8.02cm]{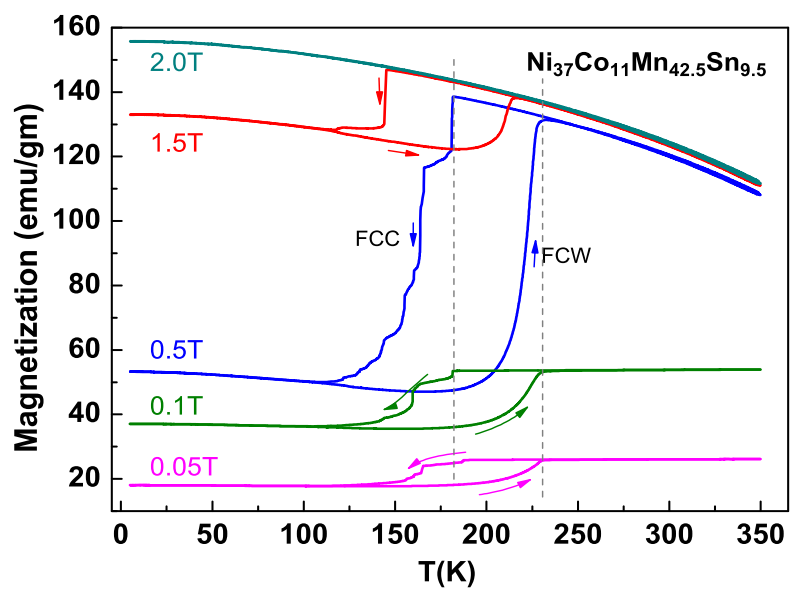}\\
  \caption{M versus T plots showing the field-cooled cooling (FCC) and field-cooled 
warming (FCW) curves in applied fields of 0.05, 0.1, 0.5, 1.5 and 2 Tesla.}\label{fig3}
\end{figure}

\Fref{fig3} shows the field cooled cooling (FCC) and field cooled warming 
(FCW) magnetization data at several different fields taken on bulk sample 
of Ni$_{37}$Co$_{11}$Mn$_{42.5}$Sn$_{9.5}$. The observed thermal hysteresis 
in the FCC and FCW curves clearly indicates that there is a first order phase 
transition (FOPT) of high magnetization austenite phase to low magnetization 
martensite phase in the cooling cycle and the reverse transformation in the 
warming cycle. The magnitude of thermal hysteresis gradually increases as 
the field is raised and has maximum hysteresis for H = 0.5 T, and by 2 T 
field, the hysteresis disappears as the austenite to martensite transformation 
is completely hindered by the field and the austenite phase gets kinetically 
arrested at low temperatures. Such behavior was seen earlier \cite{r20} in 
similar type of compounds. The austenite to martensite transformation start 
temperature (M$_{S} \sim 181$ K) in the cooling cycle and the reverse martensite 
to austenite transformation finish temperature (A$_{F} \sim 230$ K) in the 
heating cycle, are nearly same for fields up to 0.5 T. For 1.5 T, both these 
temperatures are shifted towards low temperatures to about 145 K and 214 K, 
respectively. In other words, the temperature regime over which the hysteresis 
persists is nearly the same for fields up 0.5 T, and for 1.5 T, it increases
significantly although magnitude in terms of moment has come down. Both
FCC and FCW curves merge and level off at low temperatures. The magnetization 
value at 5 K increases systematically with field as it will follow the M-H 
curve of martensite phase and also the phase fraction of austenite that 
remains arrested increases with field. 

\begin{figure}
  \centering
   \includegraphics[width=10.70cm,height=8.02cm]{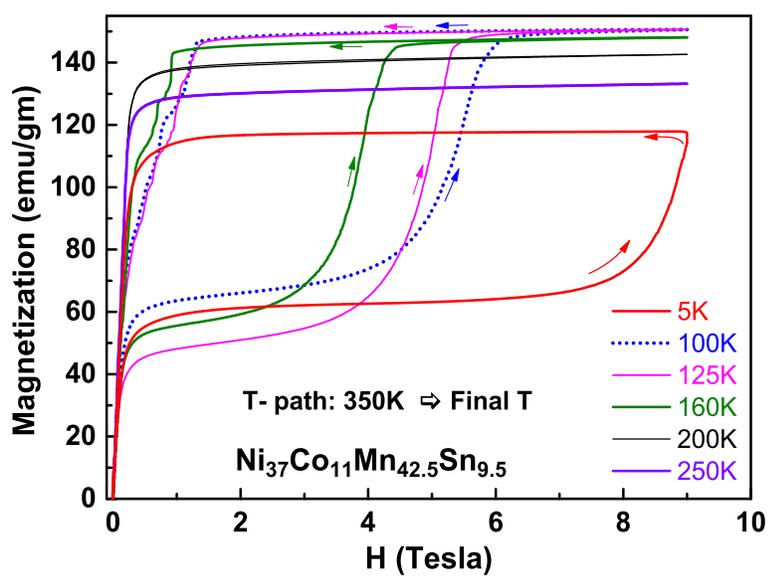}\\
  \caption{M versus H isotherms at various temperatures after cooling the sample from 
350 K.}\label{fig4}
\end{figure}

\Fref{fig4} shows the M versus H isotherms taken at different temperatures.
The final temperatures, at which M-H isotherms were recorded, are
reached by cooling the sample directly from 350 K in each case. For T $>$
200 K, the sample is in austenite phase and shows a ferromagnetic MH
with high magnetization value ($\sim 130 - 145$ emu/g). At 160 K, the sample is in
martensite phase and one observes a ferromagnetic MH isotherm with low
magnetization value ($\sim 55$ emu/g). However, with increasing field, one observes
a broad field-induced martensite to austenite transformation over a
field range of 3 T - 4 T. This trend continues as temperature is lowered
to 5 K and only the fields at which this transformation takes place go
on increasing. At 5 K, even a field of 9 T is not enough to transform
the system completely into the austenite phase.

\begin{figure}
  \centering
   \includegraphics[width=10.70cm,height=8.02cm]{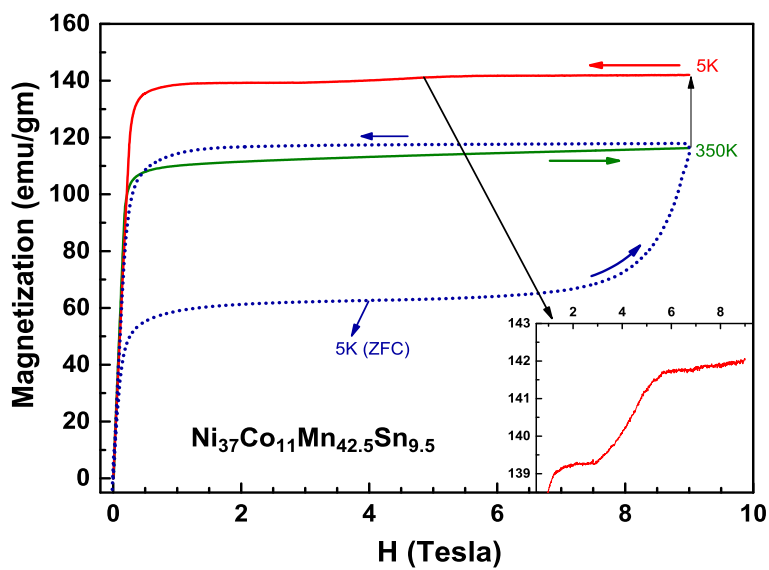}\\
  \caption{Field-cooled MH isotherms where field is raised to 9 T at 350 K (green 
curve) and sample cooled to 5 K in field after which the field is reduced to zero 
(red curve) during measurement. ZFC-MH (blue curve) at 5 K is also shown for 
comparison. Inset shows the magnified part of FC-MH at 5 K, which shows the 
devitrification of the arrested high temperature phase.}\label{fig5}
\end{figure}

\begin{figure}
  \centering
   \includegraphics[width=10.70cm,height=8.02cm]{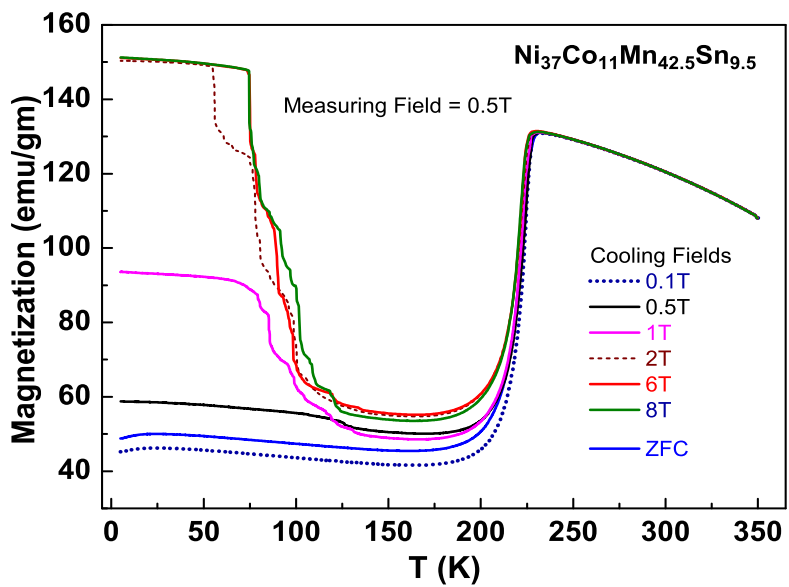}\\
  \caption{Magnetization as a function of temperature using the CHUF protocol. 
The sample is cooled in various fields and measured in 0.5 T in the warming 
cycle.}\label{fig6}
\end{figure}

In \fref{fig5}, the field is raised to 9 T at 350 K and the sample is cooled
to 5 K in field. Then, the field is isothermally reduced to zero at 5 K
(red curve). This measurement shows that some devitrification of the
arrested austenite phase takes place around 5 Tesla (inset of \fref{fig5}) as
indicated by the drop in the magnetization which is still higher than
the zero-field cooled (ZFC) magnetization at 5 K. Next, the CHUF
protocol was employed to examine the kinetic arrest of the metastable
austenite phase and its devitrification to the equilibrium martensite
phase and their phase coexistence (\fref{fig6}). The sample was cooled in
different fields ranging from 0.1 T to 8 T down to 5 K and then the
field is raised or lowered at 5 K to the value of the measuring field of
0.5 T and the measurements were carried out in the warming cycle. ZFC
magnetization curve, in which the sample was first zer-field cooled down
to 5 K and then the data were collected during the warming cycle in a 
field of 0.5 T, is also shown for
comparison. For cooling fields which are less than 0.5 T, only one
transition from a low magnetization phase to the high magnetization austenite
phase is observed which indicates that the low magnetization phase is the
equilibrium martensite phase. Whereas, for cooling fields that are
greater than 0.5 T, a sharp drop in the magnetization at low
temperatures indicates that a devitrification is taking place from a
glass-like arrested phase to an equilibrium martensite phase. On further
increasing the temperature, a reverse transformation to the high
temperature high magnetization austenite phase takes place. These observations
are discussed along with the neutron diffraction measurements performed
using the CHUF protocol in the next section.

\subsubsection{Neutron Diffraction}

\begin{figure}
  \centering
   \includegraphics[width=10.70cm,height=8.02cm]{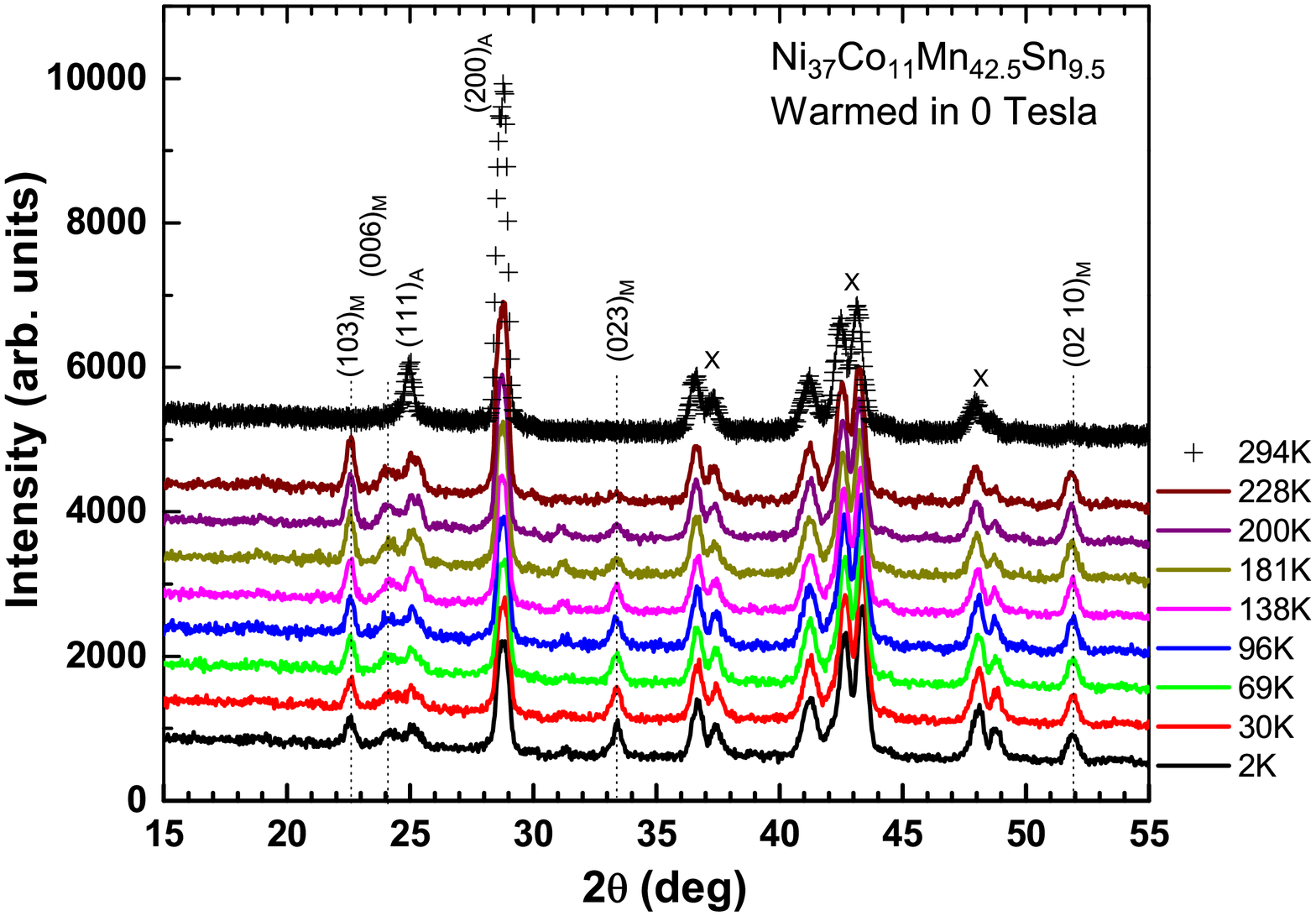}\\
  \caption{Neutron diffraction patterns as a function of temperature. Indices 
with subscript 'A' belong to the austenite phase while those with subscript 
'M' belong to the 10M martensite phase. Peaks in the regions marked as 'X' 
are contributions from the magnet shroud. Patterns have been offset for 
clarity.}\label{fig7}
\end{figure}

\begin{figure}
  \centering
   \includegraphics[width=10.70cm,height=8.02cm]{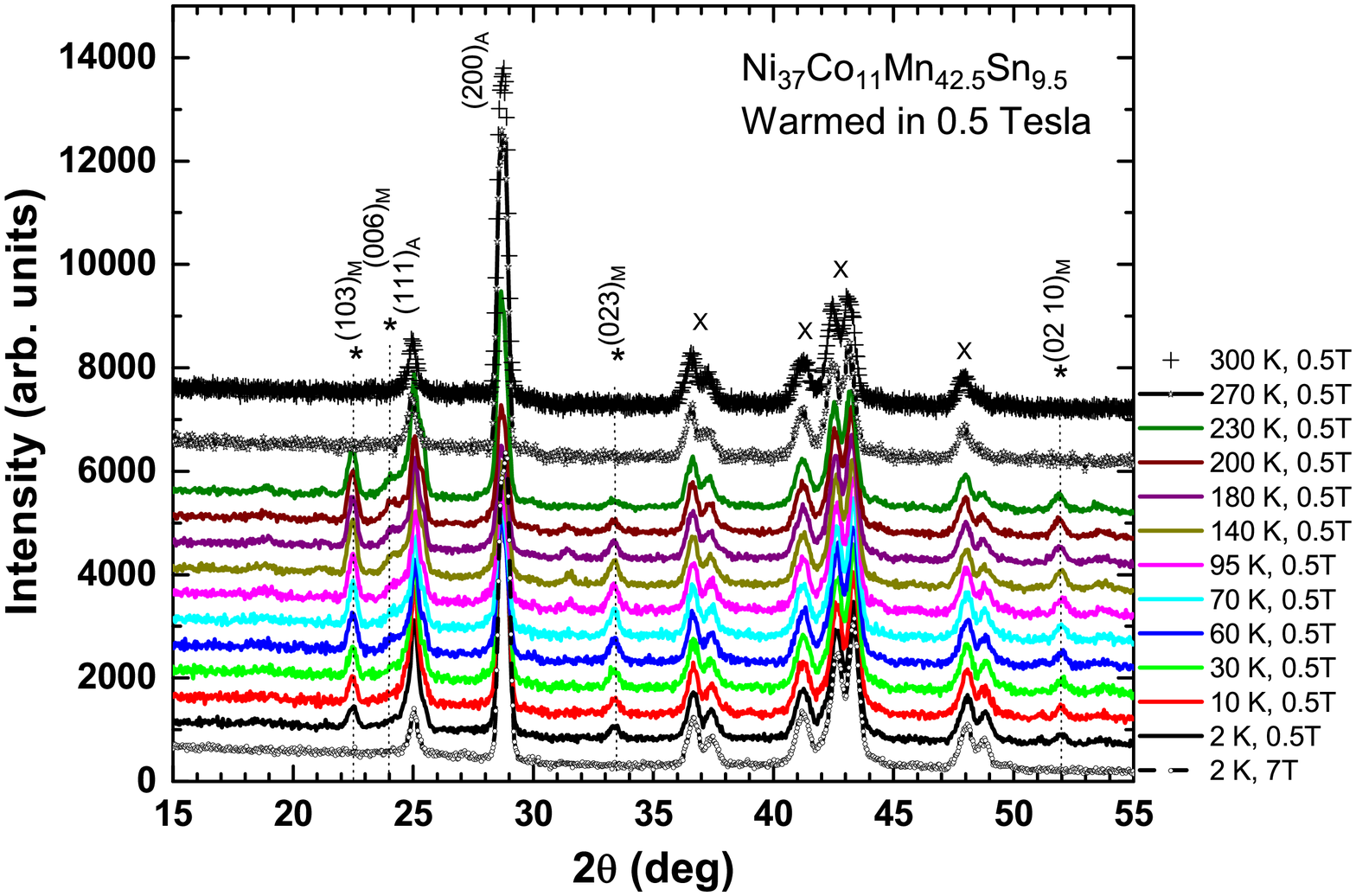}\\
  \caption{Neutron diffraction patterns obtained using the CHUF protocol. 
The sample was cooled from 300 K to 2 K in a field of 7 Tesla which is 
isothermally reduced to 0.5 T. Data were taken in the warming cycle. 
Subscripts A, M and X have the same meaning as in Fig. 7.}\label{fig8}
\end{figure}

\begin{figure}
  \centering
   \includegraphics[width=10.70cm,height=8.02cm]{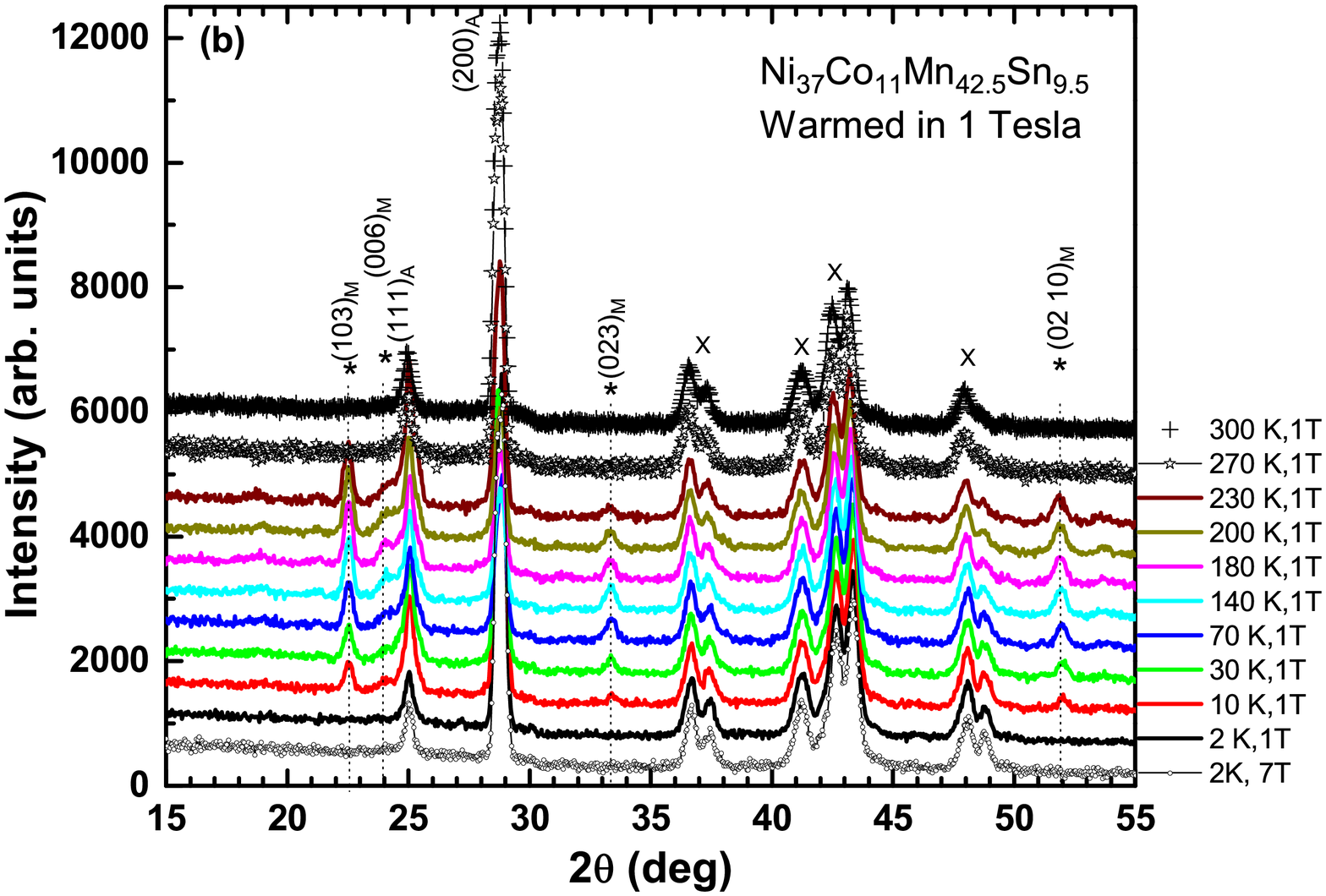}\\
  \caption{Neutron diffraction patterns obtained using the CHUF protocol. 
The sample was cooled from 300 K to 2 K in a field of 7 Tesla which is 
isothermally reduced to 1 T. Data were taken in the warming cycle. Subscripts 
A, M and X have the same meaning as in Fig. 7.}\label{fig9}
\end{figure}

Neutron diffraction patterns of bulk cylindrical specimens show that the
structure could be indexed to an L2$_{1}$-structure with cell
parameter of 5.957 {\AA} (\fref{fig7}). The bulk sample exhibited strong large
grain characteristics behaving almost like a single-crystal. The bulk
sample was therefore oriented in such a way that the (111) and (200)
reflections of the L2$_{1}$ austenite phase were strongly visible in
the ND pattern. The sample was cooled down to 2 K in this orientation
and the peaks associated with the martensite phase appear. Though it is
difficult to index the structure with the limited number of reflections,
an attempt was made using a standard indexing program and it was
observed that the martensite phase is 10M modulated with cell parameters
$a = 4.338$ {\AA}, $b = 5.534$ {\AA}  and $c = 21.21$ {\AA} and 
$\beta = 92.55^\circ$. Four reflections belonging to the martensite phase 
have been clearly identified and marked in Fig. 7. A weak martensite peak
is also observed at $\sim 31$ degrees which could be indexed as the (020)
reflection of the 10M martensite. On
warming the sample to 300 K, there is a reverse transformation to the
austenite phase in the neighborhood of 230 K. Next, a field of 7 T was
employed at 300 K and the sample was cooled again to 2 K and it was
observed that the transition to the martensite phase was completely
hindered, resulting in a kinetically arrested metastable austenite
phase as evident from the complete absence of any martensite reflections
in the ND pattern at the bottom of \fref{fig8}. Employing the CHUF protocol, the sample was cooled in a field of
7 T from 300 K several times but warmed in different fields each time.
The warming fields used were 0.5 T, 1.0 T, 1.5 T and 2 T. When the field
was reduced from 7 T to 0.5 T for the first warming cycle, it was
observed that the arrested austenite phase starts to devitrify at 2 K in
a field of 0.5 T itself (\fref{fig8}) and the expected martensite peaks
start appearing. In the next CHUF cycle, when the
field was reduced from 7 T to 1 T at 2 K, the sample remains in a
kinetically arrested state at 2 K (\fref{fig9}), as evident from the
absence of any peaks attributable to martensite phase. Devitrification
sets in at around 10 K as evidenced by the appearance of the 10M 
martensite lines mentioned above and progresses with increasing temperature up to
230 K, beyond which the system re-enters the austenite phase. With
warming fields of 1.0 T and above (figures 9-11), it is seen that the
temperature at which the devitrification sets in, increases with
increase in the warming field. It is also observed that at this higher warming
field, the intensity of the 10M martensite lines decreases implying
that the phase fraction of devitrified phase is much lower. However, it is clear from \fref{fig3}, that
cooling and warming in fields greater than 2 T would cause the
transition to be completely arrested and no devitrification would be
observed at any temperature below the standard martensite to austenite
transition temperature. However, it must be kept in mind that while
magnetization data might indicate a complete arrest of the phase, the
same data may not be able to detect the presence of a small amount of
the converted martensite phase and such a phase could be picked up by
neutron diffraction. This evident from the very low intensity lines of
the martensite phase in \fref{fig11}. It could be reasonably expected 
that there would be a complete arrest of the autenite phase at a higher
field. It is important to note here that the quantity
(H$_{c}$-H$_{w}$) which is difference between the cooling field
(H$_{c}$) and the warming field (H$_{w}$) is always positive and
in the present case, it becomes a precondition to (i) observe the
de-arrest of the metastable austenite phase which is similar to the
phenomenon of devitrification of a conventional glass upon heating and
(ii) the transformation of the devitrified equilibrium state to the high
temperature austenite phase which is analogous to melting of devitrified
glass. In \fref{fig12}, the temperature T$_{K}$ at which devitrification,
marked by the appearance of Bragg reflections attributable to the
equilibrium martensite phase, sets in, is plotted against the value of
the warming field H$_{w}$. It is observed that T$_{K}$ increases
with H$_{w}$. The set of temperature points T$_{K}$ collectively
form the kinetic arrest line which, when traversed across in the heating
cycle, would result in the de-arrest of the metastable austenite phase
to the equilibrium martensite phase. The line also indicates that at
lower warming fields, one would have a higher phase fraction of the
martensite phase when compared to higher warming fields. In order to
substantiate this, the integrated intensity of (103)$_{M}$ reflection
at 140 K, which is the temperature at which the martensite phase
fraction is expected to be at its peak value, is plotted as a function
of H$_{w}$ which is the field in which the ND patterns are recorded
in each warming cycle (\fref{fig13}). Expectedly, the martensite phase
fraction decreases at higher warming fields.

\subsection{Half-doped charge ordered compound La$_{0.5}$Ca$_{0.5}$MnO$_{3}$}
\subsubsection{DC Magnetization}

La$_{0.5}$Ca$_{0.5}$MnO$_{3}$ (LCMO) has been widely studied
and its properties have been well-documented by several authors.
Therefore, we present here only those results which are new and form the
basis for very interesting physics. The behavior of magnetization with
temperature is similar to that observed in earlier studies \cite{r12, r14, r21, r22,r23}. 
The notable features are that (i) FCC and FCW curves at
lower fields show thermal hysteresis accompanying the first order
ferromagnetic to antiferromagnetic transition, (ii) magnitude of this
thermal hysteresis gradually decreases at higher fields, culminating in
kinetic arrest of the first order phase transition and the ferromagnetic
order persists down to the lowest temperature, (iii) on application of
the CHUF protocol (\fref{fig14}), devtrification of the arrested, metastable,
ferromagnetic phase to the equilibrium antiferromagnetic state and the
subsequent reentrance to the stable ferromagnetic state at a higher
temperature is observed. \Fref{fig15} shows the magnetization curves measured
in different warming fields after cooling from 300 K in a 6 T field
along with the 6 T FCC curve. For the 0.01 T FCW curve, the kinetically
arrested ferromagnetic phase devitrifies by about 90 \% at 5 K.
Subsequently, the remaining FM phase fraction further devitrifies at $\sim
25$ K into the stable low temperature antiferromagnetic phase. On further
heating, the antiferromagnetic state reenters the higher temperature
ferromagnetic phase at $\sim 230$ K and then, finally goes into the
paramagnetic phase. This behavior unfolds more explicitly at higher
warming fields. It is also observed that the extent to which the
kinetically arrested FM phase devitrifies becomes lesser with higher
warming fields and that the temperature at which the devitrified phase
reenters the stable ferromagnetic phase decreases. This is clear from
the higher moment values at 5 K for higher warming fields. For warming
fields above 3 Tesla, the kinetically arrested phase no longer
devitrifies at any intermediate temperature.

\subsubsection{Neutron Diffraction}

Half-doped LCMO has been extensively studied using neutron diffraction \cite{r13, r24, r25,r26,r27,r28}. 
While some of the reports concentrate on determining
the magnetic and crystalline aspects of the structure at different
temperatures \cite{r13}, some deal with the local structure and
importance of the Mn-O bond lengths \cite{r26}. Magnetic field-induced
melting of the charge ordered state \cite{r27} and phase coexistence of
antiferromagnetic and ferromagnetic states due to supercooling \cite{r28}
have also been discussed. In the present neutron diffraction
measurements, we show that the FOPT from high temperature ferromagnetic
state to a low temperature antiferromagnetic state is completely or
partially arrested (depending on the path taken during cooling in a
magnetic field) due to hindered kinetics, resulting in glass-like
arrested states or GLAS. We also show that by adopting the CHUF
protocol, there is a phase coexistence of GLAS and equilibrium
(transformed) states and a de-arrest of the metastable states takes
place as one crosses the (H$_{K}$,T$_{K}$) line. \Fref{fig15} shows
the Rietveld fitted patterns of LCMO at 295 K and 9 K. The ND pattern at
295 K has been refined using the \textit{Pnma} space group. Refinement
of the ND pattern at 9 K was treated in the manner proposed by Radaelli
\etal \cite{r13} using the space group \textit{Pnma}. The
various parameters deduced from the Rietveld analysis are given in
 \tref{riet}. As expected, magnetic peaks corresponding to CE-type of ordering are
observed at 9 K. This antiferromagnetic ordering is characterized by two
Mn sublattices \cite{r24}, Mn$^{3+}$ and Mn$^{4+}$. The
propagation vectors are $[$1/2,0,1/2$]$ for Mn$^{4+}$ and
$[$1/2,0,0$]$ for Mn$^{3+}$. It is also observed that a small amount
of the room temperature phase persists at 9 K and hence, this was added
as a minority phase to the refinement along with its corresponding
ferromagnetic phase. The refinement shows that the percentage of this
minority phase is about 6 \%. Refinement of the magnetic moments shows
that they are in the \textit{a-c} plane, and $\mu_{x}$ and $\mu_{y}$
were refined separately for Mn$^{3+}$ while they were constrained to
be parallel to \textit{c}-axis for Mn$^{4+}$. The refined moment
values and the percentage phase fractions of the CE-AFM and FM phases
are given in  \tref{riet}. Next, the sample was heated to 300 K and again
cooled down to 9 K in the presence of a magnetic field of 7 Tesla. The
ND pattern at 9 K cooled in 7 Tesla field shown in \fref{fig16}, is marked by
a complete absence of any peaks signifying the CE-type AFM order seen
earlier and a strong enhancement in the intensities of the nuclear
peaks, indicating a well-defined ferromagnetic ordering. This is a clear
indication that the formation of the CE-type AFM order has been
kinetically arrested due to the field cooling of the sample in a field
of 7 Tesla. Refinement of the ND pattern at 9 K cooled in 7 T using
\textit{Pnma} space group along with a magnetic phase with
ferromagnetic order gives a moment of 4.16 $\mu_{B}$ on Mn (\tref{riet}).
This is consistent with the value obtained from magnetization
measurements. After measuring at 9 K in a cooling field of 7 T, the
field was reduced to 0.5 T (CHUF protocol), the sample was warmed from 9
K in steps. The data is plotted in a sequence in \fref{fig17}. It is seen
that up to 30 K, the diffraction spectra remain unchanged and there is
no signature of any change in the magnetic order, signifying that the
sample remains in a kinetically arrested, metastable ferromagnetic state
(magnetic GLAS) up to this temperature. At 35 K, there is an abrupt
change in the spectrum, marked by the appearance of magnetic peaks
corresponding to the CE-type AFM order (marked as asterisks). These
peaks become well-defined at 50 K and have roughly the same intensity as
the ones observed in the zero field cooled case at 9 K. Therefore, 35 K
represents the temperature around which the de-arrest or devitrification
of the metastable glassy FM phase occurs and the equilibrium AFM order
sets in. The percentage de-arrest would then be a function of the
cooling and warming fields. In this case, the refinement of the pattern
at 35 K taken during warming in a field of 0.5 T indicates that the
de-arrest is almost complete since the values of the moments on Mn$^{
3+}$ and Mn$^{4+}$ are similar to the 9 K, ZFC case (\tref{riet}). On
warming beyond 150 K, the sample re-enters a weak ferromagnetic state
before finally entering the paramagnetic state at room temperature.

In the next measurement cycle, the sample was cooled to 9 K in a field
of 7 Tesla and then, the warming field was set to 3 Tesla (CHUF
protocol). As in the earlier measurement cycle, the sample remains in a
kinetically arrested metastable ferromagnetic state at 9 K when the
external field is reduced to 3 Tesla. However, upon warming in 3 Tesla (\fref{fig18}),
the devitrification of the kinetically arrested state takes place only
around 75 K, as evidenced by the appearance of weak magnetic peaks
attributable to CE-type AFM order. In contrast to the earlier field
cycle, the intensities of the CE magnetic peaks are mcuh lower, implying
a much higher fraction of the kinetically arrested phase remaining
throughout this measurement cycle. A refinement of the pattern at 100 K
gives a moment of 3.83 $mu_{B}$ for the arrested FM phase while moments
for the devitrified CE-AFM phase come out as 2.02 $\mu_{B}$ for both Mn
$^{3+}$ and Mn$^{4+}$. The AFM peaks disappear upon warming beyond
150 K and the sample re-enters a fairly strong ferromagnetic phase at
175 K, in consonance with the magnetization measurements (\fref{fig14}).
Refinement of the pattern at 175 K of this cycle gives a moment of 3.08
$\mu_{B}$ for the re-entrant ferromagnetic phase (\tref{riet}).

It is known that the Mn-O bond lengths in LCMO at room temperature and
in the CE-AFM phase are quite different due to Jahn-Teller distortion 
\cite{r12}. Hence, it can be reasonably expected that the kinetically
arrested FM phase would have almost undistorted MnO$_{6}$ octahedra
with all the Mn-O distances being approximately similar.  \Tref{riet} shows
the Mn-O bond lengths and Mn-O-Mn bond angles at various temperature and
field combinations, and some interesting observations can be made from
the values of these quantities. The values shown for 300 K and 9 K in
zero field show the expected changes in the bond lengths and angles at
low temperature. But for the kinetically arrested FM phase at 9 K
obtained by cooling in a field of 7 T and also when the field is reduced
to 0.5 T (9 K, 0.5 T), the bond angles and lengths do not show any major
differences with the 300 K values. A direct effect of the
devitrification of the arrested phase at 35 K, 0.5 T is seen in the
Mn-O1 bond length and Mn-O1-Mn bond angle, which are comparable to the
values of the (9 K, 0 T) set. A similar behaviour is seen at 100 K in
the 3 T heating cycle. However, at 175 K warmed in 3 T field, there is a
re-entrant transition to the equilibrium FM phase, which again shows an
undistorted situation with the bond angles and lengths associated with
O1 and O2 oxygens being similar.

It is observed that the arrest temperature T$_{K}$, which has been
monitored as the temperature at which the kinetically arrested
metastable phase devitrifies, increases with the value of the warming
field H$_{w}$ as was the case with the FSMA sample and this occurs
only when the cooling field H$_{c}$ was greater than the warming
field H$_{w}$. This is consistent with the earlier observation \cite{r6} 
that the condition H$_{c} > H_{w}$ is required to observe
the devitrification whenever the FOPT temperature and the supercooling
limit T$^{*}$ fall with increase in magnetic field. To draw an
analogy, these observations are compared with the vitrification of
liquid Ge under high pressure \cite{r15} wherein it was observed that
monoatomic liquid Ge forms a glass when quenched under a pressure of
about 10 GPa. The metallic Ge glass so obtained transformed, under
ambient decompression, from a high density amorphous phase to a low
density phase. Similarly, Mishima and Stanley \cite{r16} discuss the
1-phase collapse of ice I$_{h}$ from a high-density liquid to a
high-density amorphous phase at high pressures. In the present cases,
the transition temperature drops as the magnetic field rises, analogous
to the melting point dropping with rising pressure in the case of
germanium and water. However, unlike pressure, it is possible to tune a
magnetic field effortlessly since it does not require a medium to
propagate and use it to study the microscopic nature of magnetic
field-induced transitions.

\section{Conclusions}

We have performed neutron diffraction studies on two materials belonging
to two different families of compounds, namely the magnetic shape
memory alloy Ni$_{37}$Co$_{11}$Mn$_{42.5}$Sn$_{9.5}$ and the charge-ordered
manganite La$_{0.5}$Ca$_{0.5}$MnO$_{3}$. Using magnetic field as a
useful thermodynamic variable and the CHUF protocol, we have shown 
structural evidence for the first time for (i) the kinetic arrest of the 
first order phase transition in both compounds, (ii) phase coexistence
of the metastable arrested and the equilibrium phases, and (iii) devitrification
of the metastable arrested phase.

Our diffraction studies have shown that the arrest temperature rises with 
increasing field, consistent with an earlier conjecture of Banerjee 
\etal \cite{r29} and with an extension of the regime of validity of 
the Le Chatelier Principle \cite{r30}. In cases where the
low-T equilibrium phase has a higher magnetization, transition temperature 
will rise with increasing field, and it was conjectured that the arrest 
temperature would fall with rising H. This has been confirmed through bulk 
and mesoscopic measurements in some manganites \cite{r1,r29,r30}, 
in Gd$_{5}$Ge$_{4}$ \cite{r31}, and in Ta-doped HfFe$_{2}$ \cite{r32}.
Diffraction measurements under the CHUF protocol, with H$_{w } > H_{c}$ 
would confirm the earlier conjecture of Le Chatelier Principle being valid 
during the devitrification process; some of these materials that are suitable 
for neutron diffraction studies shall be pursued.


\newpage

\begin{table}[htb]

 \caption{\label{riet} Structural parameters of La$_{0.5}$Ca$_{0.5}$MnO$_{3}$ 
obtained from Rietveld refinement of neutron  diffraction data taken at 
different temperatures and magnetic fields. Note that 0.5 T and 3 T data 
are taken in warming cycle  after the sample is cooled in 7 T field from 
300 K to 9 K and reducing the field isothermally to the respective field 
values at 9 K.}
\footnotesize\rm
\centering

\begin{tabular}{@{}cccccccc}
\br
\centre{1}{\textbf{Parameter}} & \textbf{300K, 0T} & \textbf{9K, 0T} & \textbf{9K, 7T} & \textbf{9K, 0.5T} & \textbf{35K, 0.5T} & \textbf{100K, 3T} & \textbf{175K, 3T} \\
\mr
\textit{a} &  5.4197(4) & 5.4409(4) & 5.4155(4) & 5.4142(4) & 5.4398(4) & 5.4409(4) & 5.4169(4) \\

\textit{b} &  7.6462(5) & 7.5280(5) & 7.6363(5) & 7.6354(5) & 7.5305(5) & 7.5303(5) & 7.6334(5) \\

\textit{c} &  5.4268(4) & 5.4741(4) & 5.4210(4) & 5.4202(4) & 5.4711(4) & 5.4767(4) & 5.4247(4) \\

V & 224.89(5) & 224.21(6) & 224.18(5) & 224.08(5) & 224.12(8) & 224.39(9) & 224.31(5) \\

La/Ca   x  & 0.0147(4) & 0.0228(4) & 0.0211(3) & 0.0215(5) & 0.0153(4) & 0.0162(3) & 0.0239(4) \\
  \hspace{1cm}       z  & 0.4958(3) & 0.4958(3) & 0.4964(1) & 0.4961(5) & 0.4988(1) & 0.4950(3) & 0.4982(5) \\

O1 \hspace{0.45cm} x  &  0.4931(3) & 0.4941(4) & 0.4893(3) & 0.4928(3) & 0.4907(4) & 0.5000(5) & 0.4922(3) \\
   \hspace{1cm}   z  & 0.5617(3) & 0.5660(4) & 0.5584(3) & 0.5584(3) & 0.5629(4) & 0.5705(4) & 0.5574(3) \\

O2 \hspace{0.45cm} x  & 0.2759(3) & 0.2696(4) & 0.2774(3) & 0.2751(4) & 0.2689(4) & 0.2764(4) & 0.2758(4) \\
   \hspace{1cm}   y  & 0.0319(5) & 0.0334(6) & 0.0338(5) & 0.0337(6) & 0.0332(6) & 0.0350(5) & 0.0339(5) \\
   \hspace{1cm}   z  & 0.2224(3) & 0.2275(4) & 0.2240(3) & 0.2205(4) & 0.2261(4) & 0.2212(4) & 0.2233(4) \\
B & 0.64(7) & 0.18(8) & 0.03(8) & 0.05(8) & 0.58(9) & 0.14(8) & 0.14(8) \\

Mn-O1 & 1.9410(4) & 1.9167(6) & 1.9360(4) & 1.9353(5) & 1.9145(8) & 1.9218(9) & 1.9340(4) \\
Mn-O2 & 1.9371(4) & 1.9403(6) & 1.9388(4) & 1.9270(5) & 1.9316(7) & 1.9490(8) & 1.9404(4) \\
      & 1.9505(4) & 1.9649(6) & 1.9487(4) & 1.9605(6) & 1.9722(8) & 1.9700(9) & 1.9483(5) \\

Mn-O1-Mn & 160.02(8) & 158.17(9) & 160.85(8) & 161.05(8) & 159.07(10) & 156.81(10) & 161.31(8) \\

Mn-O2-Mn & 161.09(6) & 162.37(7) & 160.50(6) & 160.36(6) & 162.34(8) & 160.08(8) & 160.56(6) \\

Magnetic order & & AFM & FM & FM & AFM+FM & AFM+FM & FM \\

AFM - Mn$^{3+}$: &&&&&&& \\
   $\mu_{x}$  ($\mu_{B}$)  &   &  1.00(6) &  &  & 1.00(6) &                  &  \\
   $\mu_{z}$  ($\mu_{B}$)  &   &  2.31(5) &  &  & 2.11(5) &  2.20(10)  &  \\
   Mn$^{4+}$: &&&&&&& \\
     $\mu_{z}$  ($\mu_{B}$)  &  &  2.46(5) &  &  & 2.26(5) &  2.20(10)  &  \\
FM - Mn: &&&&&&& \\
    $\mu_{z}$ ($\mu_{B}$) &  & 3.25(8) & 4.17(4) & 3.96 (4) & 3.73(4) & 3.83(6) & 3.08(7) \\

R$_{wp}$ & 10.7 & 19.0 & 16.1 & 19.7 & 16.5 & 25.7 & 23.4 \\

R$_{Bragg}$ & 8.9 & 10.4 & 6.8 & 7.7 & 7.6 & 11.1 & 8.1 \\

\br
\end{tabular}
\end{table}

\newpage

\begin{figure}
  \centering
   \includegraphics[width=10.70cm,height=8.02cm]{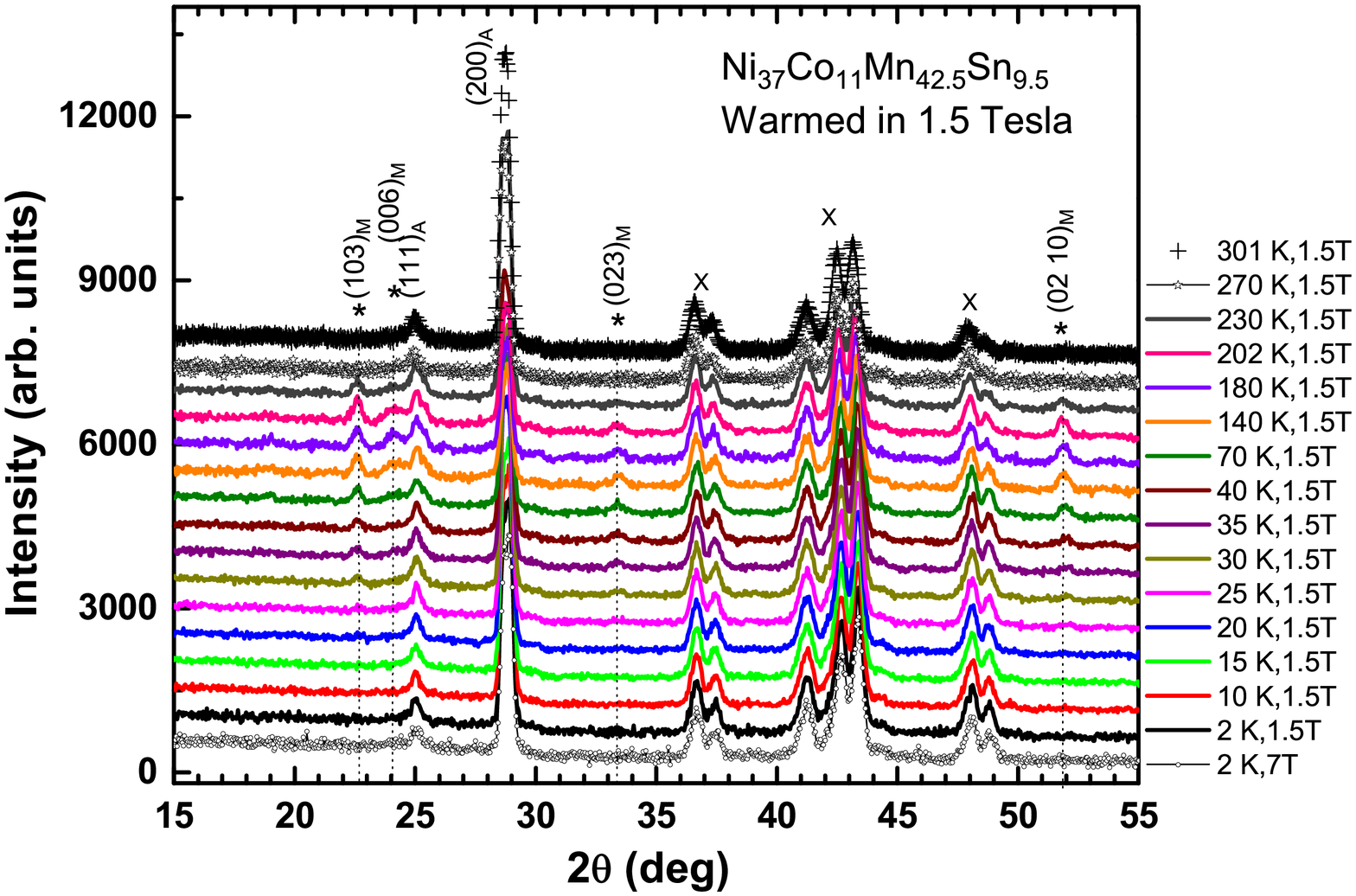}\\
  \caption{Neutron diffraction patterns obtained using the CHUF protocol. 
The sample was cooled from 300 K to 2 K in a field of 7 Tesla which is 
isothermally reduced to 1.5 T. Data were taken in the warming cycle. 
Subscripts A, M and X have the same meaning as in Fig. 7.}\label{fig10}
\end{figure}

\begin{figure}
  \centering
   \includegraphics[width=10.70cm,height=8.02cm]{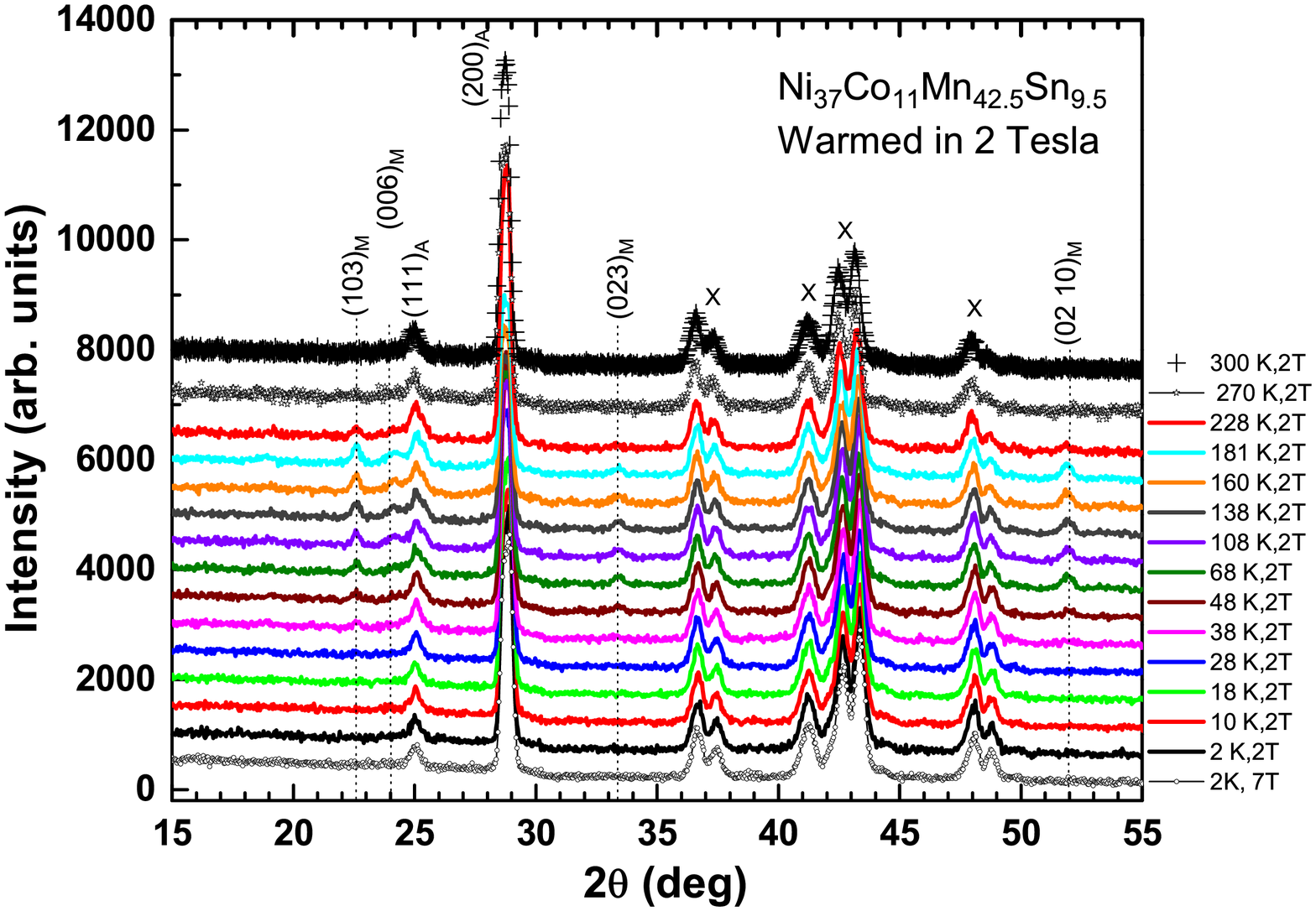}\\
  \caption{Neutron diffraction patterns obtained using the CHUF protocol. 
The sample was cooled from 300 K to 2 K in a field of 7 Tesla which is 
isothermally reduced to 2 T. Data were taken in the warming cycle. Subscripts 
A, M and X have the same meaning as in Fig. 7.}\label{fig11}
\end{figure}

\begin{figure}
  \centering
   \includegraphics[width=10.70cm,height=8.02cm]{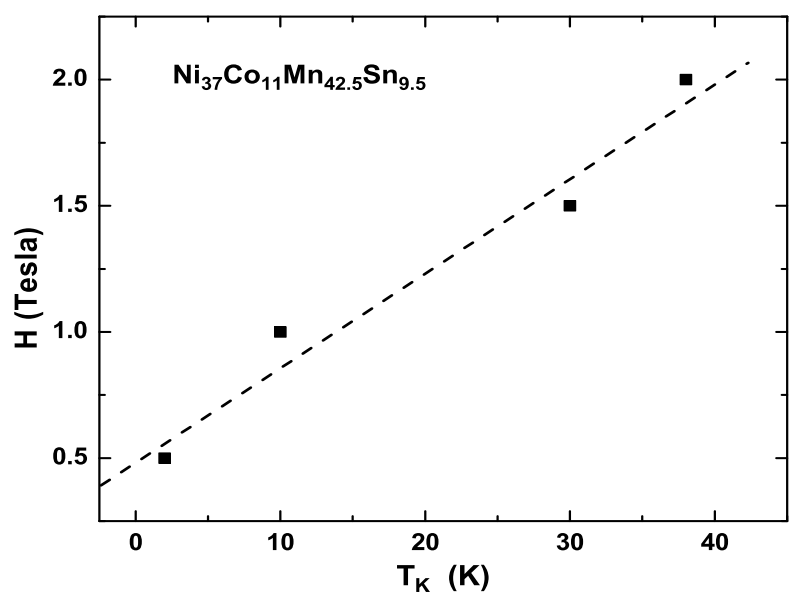}\\
  \caption{H-T diagram showing the kinetic arrest line across which there 
would be a devitrification from the arrested metastable austenite phase to 
the equilibrium martensite phase.}\label{fig12}
\end{figure}

\begin{figure}
  \centering
   \includegraphics[width=10.70cm,height=8.02cm]{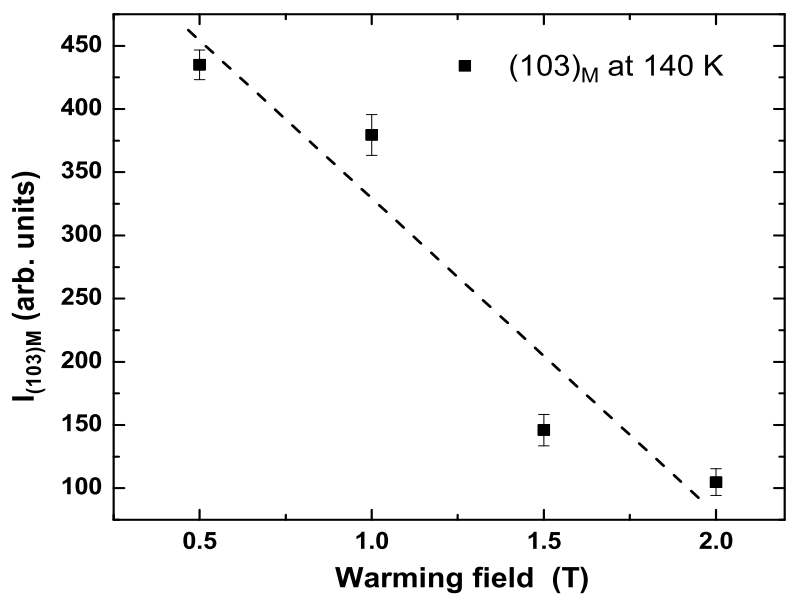}\\
  \caption{Integrated intensity of the (103) reflection of the 10M martensite 
phase (obtained from figures  9-11) as a function of the warming field at 140 
K.}\label{fig13}
\end{figure}

\begin{figure}
  \centering
   \includegraphics[width=10.70cm,height=8.02cm]{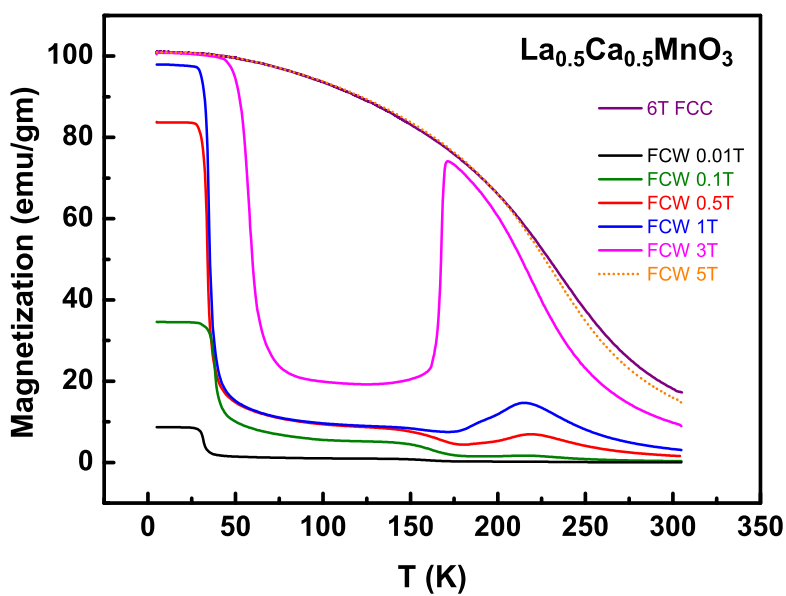}\\
  \caption{Magnetization as a function of temperature employing the CHUF 
protocol when the sample LCMO is cooled in a field of 6 T and measurements 
are carried out in different warming fields.}\label{fig14}
\end{figure}

\begin{figure}
  \centering
   \includegraphics[width=10.70cm,height=10.7cm]{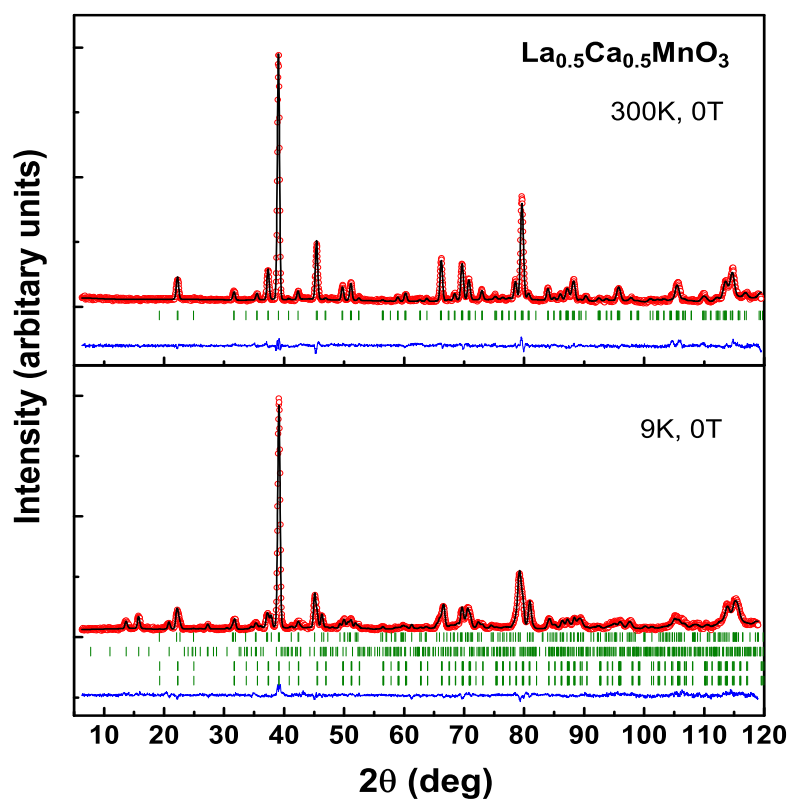}\\
  \caption{Rietveld fitted plots of LCMO at 300 K and 9 K measured in zero 
field. Observed data are filled circles and the calculated pattern is shown 
as continuous line through the data points. Difference pattern along with 
the Bragg ticks are shown below the patterns. For 9 K data, Bragg ticks 
indicate the phases (from top) majority nuclear phase, CE-AFM phase, nuclear and FM 
phases of the high temperature residual phase.}\label{fig15}
\end{figure}

\begin{figure}
  \centering
   \includegraphics[width=10.70cm,height=8.02cm]{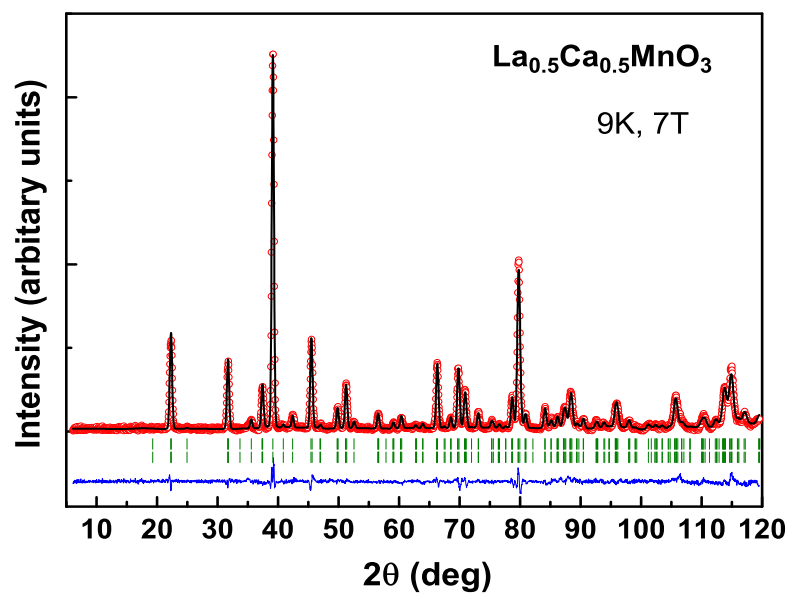}\\
  \caption{Rietveld refined pattern of the kinetically arrested metastable 
FM phase at 9 K, cooled in a field of 7 T. Bragg ticks indicate the nuclear
and FM phases.}\label{fig16}
\end{figure}

\begin{figure}
  \centering
   \includegraphics[width=10.70cm,height=8.02cm]{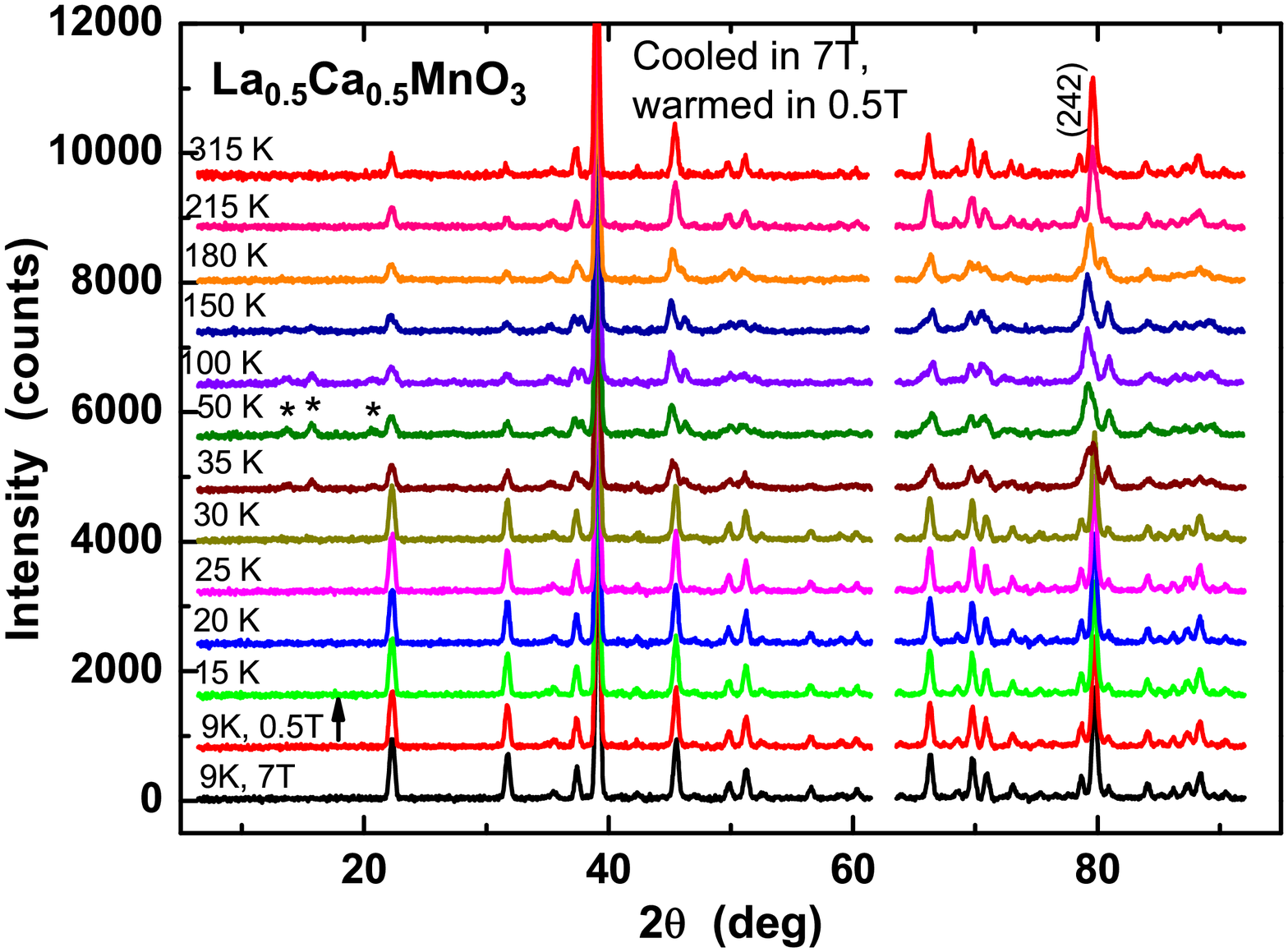}\\
  \caption{Neutron diffraction patterns obtained using the CHUF protocol. 
The sample was cooled from 300 K to 9 K in a field of 7 Tesla which is 
isothermally reduced to 0.5 T. Data were taken in the warming cycle. 
Pattern taken at 9 K, 7 T is shown at the bottom as a reference. Asterisks 
indicate the positions of the CE-AFM peaks observed upon the devitrification 
of the arrested FM phase.}\label{fig17}
\end{figure}

\begin{figure}
  \centering
   \includegraphics[width=10.70cm,height=8.02cm]{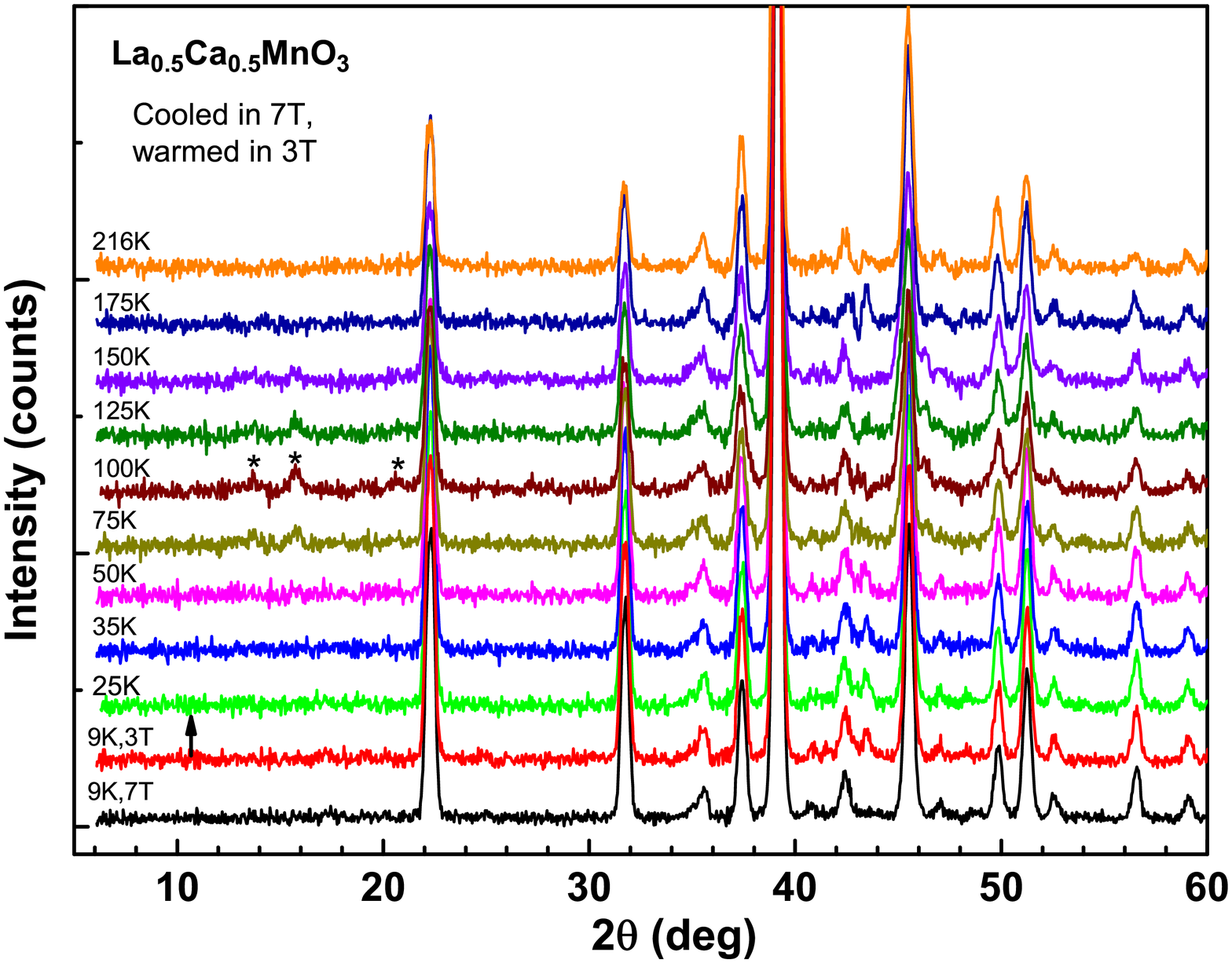}\\
  \caption{Neutron diffraction patterns obtained using the CHUF protocol. 
The sample was cooled from 300 K to 9 K in a field of 7 Tesla which is 
isothermally reduced to 3 T. Data were taken in the warming cycle. Pattern 
taken at 9 K, 7 T is shown at the bottom as a reference. Asterisks indicate 
the positions of the CE-AFM peaks observed upon the devitrification of the 
arrested FM phase.}\label{fig18}
\end{figure}

\end{document}